\documentclass[preprint,showpacs,preprintnumbers,amsmath,amssymb]{revtex4}

\usepackage{graphicx}
\usepackage{dcolumn}
\usepackage{bm}
\usepackage{color}

\usepackage{feynmf}
\usepackage{slashed}

\usepackage[utf8]{inputenc}

\begin{document}

\title{Flavor anomalies at the LHC and the R-parity violating supersymmetric model extended with vectorlike particles}

\author{Weicong Huang}
\thanks{huangwc@itp.ac.cn}
\affiliation
{State Key Laboratory of Theoretical Physics, Institute of Theoretical Physics,
Chinese Academy of Sciences, Beijing 100190, People's Republic of China
}

\author{Yi-Lei Tang}
\thanks{tangyilei15@pku.edu.cn}
\affiliation{Center for High Energy Physics, Peking University, Beijing 100871, China}

\begin{abstract}
In this paper, we consider a solution to explain the three discrepancies with the standard model (SM) predictions in flavor observables, i.e. anomalies in $B\rightarrow K^\ast\mu^+\mu^-$ and  $R_K=\mathcal{B}(B\rightarrow K\mu^+\mu^-)/\mathcal{B}(B\rightarrow K e^+e^-)$  at the LHCb and an excess in $h\rightarrow \mu\tau$ at the CMS in the context of R-parity violating (RPV) supersymmetry. We demonstrate that these anomalies can be explained within a unified framework: the minimal supersymmetry model (MSSM) extended with $5+\overline{5}$ vectorlike (VL) particles. 
The new trilinear RPV couplings involving VL particles in our model can solve the $b\rightarrow s$ anomalies , and the mixing between the SM-like Higgs boson and the VL sneutrino can yield the extra $h \rightarrow \mu \tau$ decay mode.

\end{abstract}


\maketitle
\section{Introduction}
The LHC has established the discovery of the long expected Higgs particle. So far, this boson behaves very SM-like, i.e. its dominated production and decay rates are close to the SM ones. Precision measurements of its properties would open a new window into new physics (NP) beyond the SM.  Indeed, CMS recently did observe~\cite{Khachatryan:2015kon}  a slight excess of events with a significance of $2.4\sigma$ in the lepton-flavor violating (LFV) channel $h\rightarrow \mu\tau$, which translates into a branching ratio of $\mathcal{B}(h \rightarrow \mu\tau) = (0.84^{+0.39}_{-0.37})\%$ if interpreted as a signal. Since this lepton flavor violating process is absent in the SM, various approaches have been considered to  make up the $h\tau\mu$ coupling, (For examples, see Refs.~\cite{RequireCitation5, h2mutau1, h2mutau2, h2mutau3, h2mutau4, h2mutau5, h2mutau6, h2mutau7, h2mutau8, h2mutau9, h2mutau10, h2mutau11, h2mutau12, h2mutau13, h2mutau14}), and many of them consider some types of two Higgs doublet models.

In a complementary direction, rare decays mediated by the flavor-changing neutral currents are  powerful indirect probes into NP beyond the SM. Since 2013, the LHCb collaboration has reported some anomalies in $b \rightarrow s$ transitions, including discrepancies with the SM predictions in  the angular observable $P_5^\prime$ in $B \rightarrow K^{\ast} \mu^+ \mu^-$~\cite{Aaij:2013qta} and some branching ratios~\cite{Aaij:2013aln,Aaij:2014pli}. Furthermore, an interesting hint for the lepton universality violation is observed~\cite{Aaij:2014ora} in the theoretically rather clean ratio $R_K=\mathcal{B}(B\rightarrow K\mu^+\mu^-)/\mathcal{B}(B\rightarrow K e^+e^-) = 0.745^{+0.090}_{-0.074}\pm0.036$, which departs from the SM prediction $R_K^{\mathrm{SM}}=1.0003\pm0.0001$ by $2.6\sigma$~\cite{Bobeth:2007dw}. 

It is interesting that the $b\rightarrow s$ anomalies can be explained simultaneously in a model independent approach by rather large NP contributions to the Wilson coefficients (mainly to $C_9$)~\cite{Descotes-Genon:2013wba,Altmannshofer:2013foa,Beaujean:2013soa,Hurth:2013ssa,Alonso:2014csa,Hiller:2014yaa,Ghosh:2014awa,Hurth:2014vma,Altmannshofer:2014rta, RequireCitation2}.	This has attracted considerable attention from theorists and many efforts have been made to  account for them simultaneously in one specific NP model, see for example Refs.~\cite{Hiller:2014yaa, Biswas:2014gga, Gripaios:2014tna, Varzielas:2015iva, Buras:2014fpa, Niehoff:2015bfa, Sierra:2015fma, RequireCitation3, RequireCitation4, Huang:2001me}. However, only several models  are able to address the flavor anomalies observed both at the LHCb and the CMS within a unified framework~\cite{Altmannshofer:2014cfa,Crivellin:2015mga,Crivellin:2015lwa}, and all of them utilize a $Z^\prime$ vector boson.

Supersymmetry is a well-motivated extension of the SM. However, the R-parity conserving MSSM fails to explain these anomalies simultaneously in the scenario without sources of flavor violation beyond the CKM matrix~\cite{Mahmoudi:2014mja}. Even in its more general scenario that contains flavor-changing trilinear couplings, NP effects are rather difficult to give modest contributions~\cite{Altmannshofer:2014rta}.  If R-parity is violated, the R-parity odd renormalizable Yukawa interactions of quarks and leptons with scalar superpartners  would give additional sources of flavor violation. Unfortunately, the RPV interactions in the MSSM only contribute to the operator $O^{\prime}_9$ and $O^{\prime}_{10}$~\cite{Chemtob:2004xr}. 

Introducing extra generations is one of the simplest ways to extend the SM (For a review, see \cite{4th1}. For examples, see \cite{4th2, 4th3, 4th4, 4th5, 4th6, 4th7, 4th8, 4th9, 4th10}). Compared with the extra chiral generations, the VL extensions are still viable as long as the particular vectorlike mass terms are heavy enough to escape from various experimental bounds. Supersymmetric VL extensions have also long been discussed \cite{VL1, VL2, VL3, VL4, VL5, VL6, VL7, VL8, VL9, VL10, VL11, VL12, VL13, VL14, RequireCitation1}. In order not to disturb the unification of the gauge coupling constants, which is one of the achievements of the supersymmetry \cite{SUSYGUT1, SUSYGUT2, SUSYGUT3, SUSYGUT4}, complete multiplets of the representations of the grand unification theory (GUT) $SU(5)$ group are added. Therefore, models containing copies of $5+\overline{5}$, $10+\overline{10}$ chiral superfields have been discussed in the literature. However, R-parity conserving $5+\overline{5}$ extensions of the MSSM also fails to explain the flavor anomalies. First, although the extra squarks do yield box diagrams similar to those in Refs.~\cite{Altmannshofer:2013foa, Altmannshofer:2014rta}, these contributions are suppressed due to more cross mass terms being inserted.
Second, all the charged leptons can only couple with the $H_d$, leaving us no room for a misalignment between the charged leptons' mass matrix and their Yukawa-coupling matrix unless there are large mixings between $\mu$, $\tau$ and the vectorlike leptons, which will disturb the universality of the $Zll$ vertices severely. In this paper, we consider an RPV supersymmetric model extended with one copy of the $5+\overline{5}$ vectorlike particles, and utilize it to explain all the flavor anomalies described above within a unified framework. A complete $SU(5)$ GUT model is out of our scope and left to future investigation.

The paper is organized as follows. In Sec.~\ref{BasicModel} we give a brief introduction to our RPV extension of the MSSM with $5+\overline{5}$ vectorlike particles. In Sec.~\ref{b2s} we solve the $b\rightarrow s$ anomalies utilizing the RPV operators involving vectorlike particles, and then we derive the LFV decay of the SM-like Higgs boson from our model in Sec.~\ref{h2mutau}. Finally, Sec.~\ref{Conclusions} concludes the paper.

\section{The model} \label{BasicModel}

In this  paper, we consider the MSSM extended with $5+\overline{5}$ vectorlike particles, that is to say, only $L$, $\overline{L}$, $D$, $\overline{D}$ are introduced beyond the MSSM, where $L$, $\overline{L}$ denote the leptonic $SU(2)_L$ doublets assigned with the hypercharge $-\frac{1}{2}$ and $\frac{1}{2}$ respectively and $D$, $\overline{D}$ represent the $SU(2)_L$ singlet down-type quarks assigned with the hypercharge $\frac{1}{3}$ and $-\frac{1}{3}$ respectively. We use $L_i$, $E_i$, $Q_i$, $U_i$, $D_i$, $H_u$, $H_d$ to denote the MSSM superfields, which are left-handed leptons, right-handed charged leptons, left-handed quarks, right-handed up-type quarks, right-handed down-type quarks, and the two Higgs doublet respectively, with the generation index $i$ running from 1 to 3. 

In the absence of the R-parity, gauge invariance in principle allows for baryon-number and lepton-number violating superpotential couplings. We assume that the baryon-number conserves in our model and consider only the lepton-number violating superpotential couplings involving the vectorlike particles. The pure RPV MSSM-terms are highly constrained by various experimental bounds (See \cite{Chemtob:2004xr} for a review), therefore we ignore them.

The superpotential for the $5+\overline{5}$ extension part reads
\begin{eqnarray}
W_{5+\overline{5}} = m_L \overline{L} L + m_D \overline{D} D - y^l_{i} L H_d E_i - y^d_{i} Q_i H_d D + W^{\text{RPV}}_{5+\overline{5}},
\end{eqnarray}
with
\begin{eqnarray}
W^{\text{RPV}}_{5+\overline{5}} &=& y^{QD}_{ij} Q_i L_j D + y^{L}_{ik} Q_i L D_k + y^{QD}_{i} Q_i L D + y^{UD}_{i j} U_i E_j \overline{D} \\ \notag 
&+& y^{LLE}_{ij} L_i L E_j + \epsilon_{Li} \overline{L} L_i + \epsilon_L H_u L + \epsilon_{\overline{L}} H_d \overline{L},
\end{eqnarray}
where $m_L$, $m_D$ are the vectorlike masses for the vectorlike leptons and the down-type quarks. $y^{l}_i$ and $y^{d}_i$ lead to the mixings between the SM sectors and the vectorlike sectors. $y_{ij}^{QD}$, $y^{L}_{ik}$, $y_{i}^{QD}$, $y_{ij}^{UD}$ are the corresponding coupling constants for the trilinear R-parity violating terms. $\epsilon_L$ and $\epsilon_{\overline{L}}$ yield the mixing between vectorlike leptons and the MSSM-Higgs sectors.

The corresponding supersymmetry breaking soft terms are
\begin{eqnarray}
\mathcal{L}_{\mathrm{soft}} &\supset& - m_{\tilde{L}}^2 \tilde{L}^{\dagger} \tilde{L} - m_{\tilde{\overline{L}}}^2 \tilde{\overline{L}}^{\dagger} \tilde{\overline{L}} - m_{\tilde{D}}^2 \tilde{D}^{\dagger} \tilde{D} - m_{\tilde{\overline{D}}}^2 \tilde{\overline{D}}^{\dagger} \tilde{\overline{D}} + [ - A^l y_{i}^l \tilde{L} H_d \tilde{E_i} - A^d y_{i}^d \tilde{Q} H_d \tilde{D} \nonumber \\
&+& A^{QD} (y^{QD}_{ij} \tilde{Q}_i \tilde{L_j} \tilde{D} + y^{QD}_i \tilde{Q}_i \tilde{L} \tilde{D}) + A^{UD} y_{ij}^{UD} \tilde{U}_i \tilde{E}_j \tilde{\overline{D}} + A^{LLE} y^{LLE}_{ij} \tilde{L}_i \tilde{L} \tilde{E} \nonumber \\
&+& B_{\overline{L} L} \epsilon_{L i} \tilde{\overline{L}} \tilde{L}_i + B_L \epsilon_L H_u \tilde{L} + B_{\overline{L}} \epsilon_{\overline{L}} H_d \tilde{\overline{L}} + m_{dL}^2 H_d^{\dagger} \tilde{L}  + m_{uL}^2 H_u^{\dagger} \tilde{\overline{L}}  + \text{H.c.} ]
\end{eqnarray}

For later convenience, we write down the MSSM superpotential
\begin{eqnarray}
W_{\mathrm{MSSM}} = W_{\mathrm{MSSM}}^{\mathrm{RPC}} =  \mu H_u H_d + y^u_{ij} Q_i H_u U_j - y^d_{ij} Q_i H_d D_j - y^l_{ij} L_i H_d E_j ,
\end{eqnarray}
and the corresponding soft terms
\begin{eqnarray}
\mathcal{L}_{\mathrm{soft}} &\supset& - m^{Q2}_{ij} \tilde{Q}^{\dagger}_i \tilde{Q}_j - m^{U2}_{ij} \tilde{U}^{\dagger}_i \tilde{U}_j - m^{D2}_{ij} \tilde{D}^{\dagger}_i \tilde{D}_j - m^{L2}_{ij} \tilde{L}^{\dagger}_i \tilde{L}_j - m^{E2}_{ij} \tilde{E}^{\dagger}_i \tilde{E}_j \nonumber \\
&-& m_{H_u}^2 |H_u|^2 - m_{H_d}^2 |H_d|^2 + (A^u_{ij} \tilde{Q}_i H_u \tilde{U}_j - A^d_{ij} \tilde{Q}_i H_d \tilde{D}_j - A^l_{ij}\tilde{L}_i H_d \tilde{E}_j   \nonumber \\
&-&  B \mu H_u^{\dagger} H_d + \mathrm{H.c.} )\, .
\end{eqnarray}

At the end of this section, we comment on the lepton-number violating interactions due to RPV, which are tightly constrained  by the experimental bounds, e.g., the neutrinoless double beta decay or the neutrino masses.  The RPV induced neutrinoless double beta decay usually requires
$\nu \tilde{q} q (Q_iL_jD_{(k)})$  and  $u e^{\pm} \tilde{q}$ ($Q_1 L_1D_{(i)}$ or $U_1 E_1 \overline{D}$) vertices, or large mixture between neutrinos and neutrilinos (For a review, see \cite{Chemtob:2004xr}. For examples, see \cite{RPV0vee1, RPV0vee2}). The neutrino masses can be induced from the vacuum expectation value of the sneutrinos at tree-level or might come from the quark/squark loops induced by the $Q_iL_jD_{(k)}$ vertices. However, in the following text we will see that we only need the $U_3 E_{1,2} \overline{D}$, $L_2 L E_3$ and $L_3 L E_2$ vertices. All the other trilinear RPV coupling constants can always be set small enough in order to avoid the unwanted vertices mentioned before. We can also adjust the parameter in order to forbid the mixture between MSSM (s)leptons and the VL or Higgs sectors, thus vacuum expectation values of the MSSM sneutrinos can be avoided.

\section{Explaining the $b \rightarrow s $ Anomalies} \label{b2s}
The effective Hamiltonian for $b \rightarrow s $ transitions can be written as
\begin{eqnarray}
\mathcal{H}_{\mathrm{eff}} \supset -\frac{4 G_F}{\sqrt{2}} V_{tb} V_{ts}^* \frac{e^2}{16 \pi^2} \sum_i(C_i O_i + C_i^{\prime} O_i^{\prime}) + \mathrm{H.c.}\ ,\label{btosHamiltonian}
\end{eqnarray}
where $V_{ij}$ denotes the CKM matrix elements and $C_i^{(\prime)}$ are the Wilson coefficients of the effective operators $O_i^{(\prime)}$. 
According to the global fits~\cite{Altmannshofer:2014rta}, we consider new physics effects in the following set of operators, 
\begin{eqnarray}
O_9^\mu = \left(\overline{s} \gamma_{\mu} P_{L} b\right)\left(\overline{\mu} \gamma^{\mu} \mu \right),\quad
O_{10}^\mu = \left(\overline{s} \gamma_{\mu} P_{L} b \right) \left(\overline{\mu} \gamma^{\mu} \gamma^5 \mu \right), 
\end{eqnarray}
which is one of the best fit scenarios with 
\begin{eqnarray}
-1.6\,(-1.4)&<&\mathrm{Re}\, C_9^{\mu,\mathrm{NP}} <(-0.6)\,-0.3 ,\notag\\
-0.4\,(-0.2)&<&\mathrm{Re}\, C_{10}^{\mu,\mathrm{NP}} <(0.5)\,0.8 ,\label{GFit}
\end{eqnarray}
at ($1 \sigma$) $2 \sigma$ level.
Besides, the rest of operators involving muons, electrons and taus are perfectly compatible with the SM expectations in the fitting.

In the R-parity conserving MSSM, 
the only way to break the $e-\mu$ universality is through box diagrams involving light smuons while selections are decoupled. In such case, non-negligible contributions to $C_9^\mu$ and $C_{10}^\mu$ can only come from the boxes induced by the flavor violation in the squark soft masses. In Ref.~\cite{Altmannshofer:2013foa}, rather modest contributions to $C_9^\mu$ and $C_{10}^\mu$ of $\gtrsim 0.5$ is obtained with an extremely light spectrum which is strongly disfavored by the direct searches. Here we make a more conservative estimation of the contributions according to the bounds from LHC. Assuming maximal mixing of the left-handed bottom and strange squarks, the wino boxes contributions (dominate over those from bino and mixed wino-bino boxes) read~\cite{Altmannshofer:2013foa}
\begin{equation}
\left(V_{ts}^\ast V_{tb}\right)C_9^{\mathrm{box}} \simeq \frac{1}{s_W^2} \frac{5}{192}\frac{m_W^2}{m_{\tilde{d}}^2}\, (\delta_{bs}^L)\, f_9^{\mathrm{box}}\left(\frac{m_{\tilde{l}}^2}{m_{\tilde{d}}^2},\frac{m_{\tilde{W}}^2}{m_{\tilde{d}}^2}\right)
\end{equation}
with the loop function $f_9^{\mathrm{box}} $ given in the Appendix~\ref{app_box}. For $m_{\tilde{l}}=m_{\tilde{W}} = 130$ GeV, $m_{\tilde{d}} = 800$ GeV and $\delta_{bs}^L = -0.4$, we obtain contributions $C_9^{\mathrm{box}} = -C_{10}^{\mathrm{box}} = -0.2$. Obviously, these box contributions are insufficient to account for the anomalies.

In the R-parity violating supersymmetry models, there are extra tree-level sources of flavor violation from the trilinear R-parity violating terms. Unfortunately,  such trilinear terms involving pure MSSM superfields can never yield the effective operators $O_{9,10}$ which only involve the left-handed down-type quarks since the charged-leptons always couple to the right-handed down-type quarks in $Q_i L_j D_k$ vertices. In our model extended with one copy of the $5+\overline{5}$ vectorlike superfields, this problem can be solved with the trilinear couplings $U_i E_j \overline{D}$, i.e.  left-handed down-type quarks  in the SM couple to charged leptons by their mixing with the vectorlike quarks. 

Integrating out the squarks, we obtain the following effective Hamiltonian for $b \rightarrow s\mu\mu$ transitions
\begin{eqnarray}
\mathcal{H}_{\mathrm{eff}} &\supset& - V_{Db} V_{Ds}^\ast \sum_{k=1}^3  \frac{y_{k 2}^{UD *} y_{k 2}^{UD} }{ m_{\tilde{u}_{Rk}}^2} (\overline{s} P_R\, \mu ) ( \overline{\mu} P_L b ) + \mathrm{H.c.}\label{BeforeFierz} \notag\\
 &=&  - V_{Db} V_{Ds}^* \sum_{k=1}^3   \frac{y_{k 2}^{UD *} y_{k 2}^{UD}}{2 m_{\tilde{u}_{Rk}}^2} (\overline{s} \gamma^{\mu} P_L b)( \overline{\mu} \gamma_{\mu} P_R\, \mu )+ \mathrm{H.c.}\  ,\label{AfterFierz}
\end{eqnarray}
where $m_{\tilde{u}_{Rk}}$ is the mass of the kth right-handed up-type squark, $V_{Db}, V_{Ds}$ denote elements of the extended CKM matrix in our model and the Fierz transformation is applied in the second line. Notice that
\begin{eqnarray}
V_{Ds} \approx - \frac{y_2^d v \cos \beta}{m_D}, \nonumber \\
V_{Db} \approx - \frac{y_3^d v \cos \beta}{m_D}, \label{V_Formula}
\end{eqnarray}
where $v = \sqrt{v_u^2 + v_d^2} = 174 \text{ GeV}$ is the electroweak vacuum expectation value, while the $\tan \beta = \frac{v_u}{v_d}$  is the ratio of the vacuum expectation values of the $H_u^0$ and $H_d^0$.
Immediately, the Wilson coefficients in term of the R-parity operators read
\begin{eqnarray}
C_9^{\mu,\mathrm{VLRPV}} = C_{10}^{\mu,\mathrm{VLRPV}} = \frac{\sqrt{2} \pi^2 V_{Db} V_{Ds}^*}{G_F V_{tb} V_{ts}^* e^2} \sum_{k=1}^3 \frac{y_{k2}^{UD *} y_{k2}^{UD}}{m_{\tilde{u}_{Rk}^2}}\ .\label{RPVWC}
\end{eqnarray}

The magnitude of $C_9^{\mu,\mathrm{VLRPV}} $ and $C_{10}^{\mu,\mathrm{VLRPV}}$ is related with the mixing parameters $V_{Db}$ and $V_{Ds}$, which are mainly constrained by the unitarity of the extended CKM matrix, i.e.
\begin{eqnarray}
|V_{us}|^2 + |V_{cs}|^2 + |V_{ts}|^2 + |V_{Ds}|^2 &=& 1\ , \nonumber \\
|V_{ub}|^2 + |V_{cb}|^2 + |V_{tb}|^2 + |V_{Db}|^2 &=& 1\ . \label{UnitaryOfThisModel}
\end{eqnarray}
According to the data from the PDG~\cite{Agashe:2014kda},
\begin{eqnarray}
|V_{us}|^2 + |V_{cs}|^2 + |V_{ts}|^2 &=& 1.025 \pm 0.032 \ ,\notag \\
|V_{ub}|^2 + |V_{cb}|^2 + |V_{tb}|^2 &=& 1.042 \pm 0.065 \label{CKMUnitary}\ .
\end{eqnarray}
The data together with the error bar we adopt come from the direct measurements of the meson behaviors without any fittings using the $3 \times 3$ unitary properties. In the RPV cases the trillinear vertices together with the sparticle propagators might fake the W-boson induced effects and then disturb the semileptonic decay of the mesons, thus the values of the measured CKM matrix elements are deviated. However, just like what has been mentioned in Sec.~\ref{BasicModel}, all the unwanted terms (mainly $Q L_i D$-like terms) can be turned down in order to avoid these problems. Therefore, in this paper, we ignore these effects. Note the error bar in the second line of (\ref{CKMUnitary}) is mainly controlled by the uncertainty in $|V_{tb}|$. Hence, we obtain upper bounds on the mixing parameters, $|V_{Ds}| \lesssim 0.084$ and $|V_{Db}| \lesssim 0.15$. Assuming $m_{\tilde{u}, \tilde{c}} \gg m_{\tilde{t}}$ and plugging in $G_F = 1.1663787\times10^{-5} \mathrm{GeV}^{-2}$, $\alpha_e(m_b)=1/133$, $|V_{ts}|=0.0404$ and  $|V_{tb}|=1.021$, the contributions (\ref{RPVWC}) become
\begin{eqnarray}
C_9^{\mu,\mathrm{VLRPV}} = C_{10}^{\mu,\mathrm{VLRPV}} = \frac{|y_{32}^{UD}|^2 V_{Db} V_{Ds}^\ast\ }{m_{\tilde{t}_{R}^2}}(18 \mathrm{TeV})^2.
\end{eqnarray}
Here we give a benchmark point that solves the anomalies. Take $m_{\tilde{t}} = 900$ GeV, $y_{32}^{UD} = 0.4$, $V_{Db} = 0.1$ and $V_{Ds} = -0.05$, then $C_9^{\mu,\mathrm{VLRPV}}= C_{10}^{\mu,\mathrm{VLRPV}}=-0.3$. Note that in this case, given the condition $|y_{2,3}^d|<1$ that the perturbative theory is available, from (\ref{V_Formula}) we can see that the mass term $m_D$ is constrained as
\begin{eqnarray}
|m_D| \lesssim \frac{ |y_{2,3}^d| v \cos \beta}{|V_{Ds}|}. 
\end{eqnarray}
Assuming $\tan \beta = 2$ results in $|m_D| \lesssim 800 \text{ GeV}$, which is compatible with the experimental data (See \cite{Agashe:2014kda, bprime1, bprime2}).
 
Combining these with the MSSM box contributions $C_9^{\mathrm{box}} = -C_{10}^{\mathrm{box}} \sim 0.2$, we finally get the total contributions of new physics $C_9^{\mathrm{NP}} = -0.5$ and $C_{10}^{\mathrm{NP}} = -0.1$, which is compatible with the global fitting results (\ref{GFit}) at $2\sigma$ level.

The $B_s$ meson mixing and the rare muonic decays of the neutral B mesons provide important constraints on NP scenarios,  and the effective Hamiltonian~(\ref{btosHamiltonian}) for the $b\rightarrow s$ transitions is relevant for these processes. As for the $B_s$ meson mixing, since there is no tree-level contribution in our model, and we expect the one-loop box diagrams to be highly suppressed by the heavy squark mass ($\sim 800 \text{ GeV}$) and the small mixings between the VL quarks $D$, $\overline{D}$ and the SM ones, so we neglect this constraint and only consider the latter one. The amplitude for the $B_s \rightarrow \mu^+\mu^-$ decay is dominated by the axial vector operator $O_{10}^{\mu}$ while the vector contribution from $O_{9}^{\mu}$ vanishes, and thus the branch ratio in our model can be well approximated by
\begin{equation}
\mathcal{B}(B_s\rightarrow\mu^+\mu^-)=\left|\frac{C_{10}^{\mu,\mathrm{SM}}+C_{10}^{\mu,\mathrm{NP}}}{C_{10}^{\mu,\mathrm{SM}}}\right|^2 \mathcal{B}(B_s\rightarrow\mu^+\mu^-)_{\mathrm{SM}}
\end{equation}
with the SM prediction $\mathcal{B}(B_s\rightarrow\mu^+\mu^-)_{\mathrm{SM}} = (3.65\pm 0.23)\times 10^{-9}$\cite{Bobeth:2013uxa}.
Recently this rare decay has been observed from the combined analysis of CMS and LHCb data~\cite{CMS:2014xfa} with a branch ratio of $\mathcal{B}(B_s \rightarrow \mu^+\mu^-)= (2.8^{+0.7}_{-0.6})\times 10^{-9}$, which translates into~\cite{Sierra:2015fma} $-0.25 < C_{10}^{\mu,\mathrm{NP}}/C_{10}^{\mu,\mathrm{SM}}<0.03$ (at the $1\sigma$ level). For the benchmark point given above, we have $ C_{10}^{\mu,\mathrm{NP}}/C_{10}^{\mu,\mathrm{SM}} = 0.025$ and hence compatible with experimental measurement at $1\sigma$ level.

\section{Explaining the Higgs decay $h \rightarrow \mu \tau$} \label{h2mutau}
The effective operators describing the $h \rightarrow \mu \tau$ decays are given by
\begin{eqnarray}
\mathcal{L} \supset  - y_{\mu \tau} \overline{\mu}_L \tau_R h  - y_{\tau \mu} \overline{\tau}_L \mu_R h+ \mathrm{H.c.} \ ,
\end{eqnarray}
yielding the branching ratio
\begin{eqnarray}
\mathcal{B}(h \rightarrow \mu\tau) \simeq \frac{m_h}{8\pi \Gamma_{\mathrm{SM}}}(|y_{\mu\tau}|^2+|y_{\tau\mu}|^2)
\end{eqnarray}
where $\Gamma_{\mathrm{SM}}\simeq 4.1$ MeV is the decay width for a 125 GeV Higgs in the SM~\cite{Dittmaier:2012vm}. Correspondingly, the expected values of the effective couplings to explain the experimental results are
\begin{equation}
\sqrt{|y_{\mu\tau}|^2+|y_{\tau\mu}|^2}\simeq 0.0026\pm 0.0006\, . \label{effcoupval}
\end{equation}

In the MSSM, the Yukawa coupling matrix of the charged lepton to the SM-like Higgs is always proportional to their mass matrix and thus there is no $h \mu \tau$ vertices after rotating the charged lepton sectors into their mass eigenstates. It is interesting that sneutrinos share the same quantum numbers with the neutral Higgs fields, so they can mix with the Higgs boson and then produce lepton flavor violating Higgs couplings in the RPV supersymmetric models. As mentioned above, the mixings between the MSSM sneutrinos and the Higgs boson usually result in too heavy SM-like neutrinos (See \cite{Chemtob:2004xr, Barbier:2004ez} for discussions), we have ignored the relevant R-parity violating terms at the begin of our model building. Then the ``Higgs" superpotential relevant to the electroweak symmetry breaking in our model is  
\begin{eqnarray}
W_{\mathrm{Higgs}} = \mu H_u H_d + m_L \overline{L} L + \epsilon_L H_u L + \epsilon_{\overline{L}} H_d \overline{L}\ ,
\end{eqnarray}
which yields the following Higgs potential for the neutral scalar fields $H_u^0, H_d^0, \tilde{L}^0$ and $\tilde{\overline{L}}\,^0$ :
\begin{eqnarray}
V_{\mathrm{eff}}(H_u^0, H_d^0, \tilde{L}^0,\tilde{\overline{L}}\,^0) &=& \frac{g_1^2+g_2^2}{8} ( |H_u^0|^2 - |H_d^0|^2 - |\tilde{L}^0|^2 + |\tilde{\overline{L}}\,^0 |^2)^2  
+ |\epsilon_{\overline{L}} H_d^0 - m_L \tilde{L}^0|^2 \notag \\ 
&+& |\epsilon_{L} H_u^0 + m_L \tilde{\overline{L}}\,^0 |^2 
+ |\mu H_d^0 + \epsilon_L \tilde{L}^0|^2 + |\mu H_u^0 - \epsilon_{\overline{L}}\tilde{\overline{L}}\,^0|^2 \notag \\
&+& m_{H_u}^2 |H_u^0|^2 + m_{H_d}^2 |H_d^0|^2 + m_{\tilde{L}}^2 |\tilde{L}^0|^2 + m_{\tilde{\overline{L}}}^2 |\tilde{\overline{L}}\,^0|^2  \nonumber \\
&+& ( B \mu H_u^0 H_d^0 + B_L \epsilon_L H_u^0 \tilde{L}^0 + m_{dL}^2 H_d^{0 \ast} \tilde{L}^0 + m_{uL}^2 H_u^{0 \ast} \tilde{\overline{L}}\,^0 + \mathrm{H.c.} ) \ . \label{EffPotentialTree}
\end{eqnarray}
After the electroweak spontaneously symmetry breaking, all these scalar fields might acquire vevs. Note that the mixing between the $H_u^0$ and $\tilde{\overline{L}}\,^0$ might bother the properties of the SM-like Higgs boson severely, we decouple $\tilde{\overline{L}}\,^0$ by assuming the soft mass $m^2_{\tilde{\overline{L}}} \gg m^2_{H_u}$. for simplicity. In addition, we could always redefine the $H_d$ and $\tilde{L}$ field by a rotation so that $v_L  \equiv \langle\tilde{L}\rangle = 0$. Taking account of these, the minimization conditions are given by
\begin{eqnarray}
M_{H_d}^2 &=& \frac{2 \lambda_{\mathrm{eff}} v_d (v_u^2 - v_d^2) - B \mu v_u}{v_d}, \nonumber \\
M_{H_u}^2 &=& \frac{ - 2 \lambda_{\mathrm{eff}} v_u (v_u^2 - v_d^2) - B \mu v_d}{v_u}, \nonumber \\
M_{dL}^2 &=& - B_L \epsilon_L \frac{v_u}{v_d}. \label{TadpoleResults}
\end{eqnarray}
where $v_u = \langle H_u^0\rangle, v_d = \langle H_d^0\rangle$ and we have used a set of shorthand notations for convenience: $M_{H_u}^2 = |\mu|^2 + m_{H_u}^2 + \epsilon_L^2$, $M_{H_d}^2 = |\mu|^2 + m_{H_d}^2 + \epsilon_{\overline{L}}^2$, $M_{\tilde{L}}^2 = m^2_L + m_{\tilde{L}}^2 + \epsilon_L^2$, $M^2_{dL} = m^2_{dL} - \epsilon_{\overline{L}} m_L + \mu \epsilon_L $ and
$\lambda_{\mathrm{eff}} = \frac{g_1^2 + g_2^2}{8}$.

The tree-level Higgs mass-squared matrix can be calculated from the potential (\ref{EffPotentialTree}) and read in the basis $\frac{1}{\sqrt{2}}\mathrm{Re}(H_u^0, H_d^0, \tilde{L}^0 )$ after substituting conditions (\ref{TadpoleResults})
\begin{eqnarray}
M^2_{\text{Higgs+$\tilde{L}$}} = \left[
\small{
\begin{array}{ccc}
4 \sin^2 \beta v^2 \lambda_{\mathrm{eff}} - B \mu \cot \beta & -4 \sin \beta \cos \beta v^2 \lambda_{\mathrm{eff}} + B \mu & B_L \epsilon_L \\
-4 \sin \beta \cos \beta v^2 \lambda_{\mathrm{eff}} + B \mu & 4 \cos^2 \beta v^2 \lambda_{\mathrm{eff}} - B \mu \tan \beta & -B_L \epsilon_L \tan \beta \\
B_L \epsilon_L &  -B_L \epsilon_L \tan \beta & M_{\tilde{L}}^2 + 2 v^2 ( \cos^2 \beta - \sin^2 \beta ) \lambda_{\mathrm{eff}} 
\end{array}}
\right], \label{MassMatrixBeforeRotation}
\end{eqnarray}
with $v^2=v_u^2 + v_d^2 \simeq(174 \mathrm{GeV})^2$ and $\tan \beta = \frac{v_u}{v_d}$. The SM-like Higgs is one of the mass eigenstates diagonalizing the above matrix and can be parametrized as
\begin{eqnarray}
h = \ V_{u}\, h_u^0 + V_{d}\, h_d^0 + V_{l}\, \tilde{l}^0 \ , 
\end{eqnarray}
where $(h_u^0,\, h_d^0,\, \tilde{l}^0 )= \frac{1}{\sqrt{2}} \mathrm{Re}(H_u^0, H_d^0, \tilde{L}^0) $ and $|V_{u}|^2 + |V_{d}|^2 + |V_{l}|^2 = 1$. The effective couplings $y_{\mu\tau}$ and $ y_{\tau\mu}$ are thereby
\begin{eqnarray}
y_{\mu\tau} = V_l y_{23}^{LLE}\ , \quad y_{\tau\mu} = V_l y_{32}^{LLE}
\end{eqnarray}

In order to get the mixing coefficient $V_l$, we need to diagonalize the squared mass matrix (\ref{MassMatrixBeforeRotation}). Note that the mixings between the Higgs and sneutrino  are controlled by $B_L \epsilon_L$, we can make a perturbative diagonalizing with respect to $(B_L\epsilon_L \sec \beta) /M_{\tilde{L}}^2 \ll 1$(see detail in Appendix B). At the lowest order, we obtain 
\begin{eqnarray}
V_{l}^{\mathrm{lowest}} = \frac{2 B_L \epsilon_L \lambda_{\mathrm{eff}} v^2 \sin \beta \sin 4 \beta}{ M_{\tilde{L}}^2 (B \mu +  (\sin 6\beta - \sin 2\beta)\lambda_{\mathrm{eff}}v^2)}. \label{VlLowest}
\end{eqnarray}
For a moderate $(B_L\epsilon_L \sec \beta) /M_{\tilde{L}}^2$, we must include higher order contributions or diagonalize (\ref{MassMatrixBeforeRotation}) numerically .  
Here a sample point is given to explain the decay $h \rightarrow \mu \tau$: $\tan \beta = 2$, $M_{\tilde{L}} = 400$ GeV, $B \mu = -(3 \times 10^2 \,\mathrm{ GeV})^2$ and $B_L = \epsilon_L = 170$ GeV yield $V_{l}^{\mathrm{lowest}} = 0.0049$ while $V_{l} = 0.0057$ in numerical calculations. Correspondingly, Eq.(\ref{effcoupval}) holds if only $\sqrt{(y_{23}^{LLE})^2 + (y_{32}^{LLE})^2} \simeq 0.46$. 

According to the work of Ref.~\cite{Harnik:2012pb}, the effective couplings (\ref{effcoupval}) for a very SM-like Higgs are compatible with other relevant favor constraints, e.g., from $\tau \rightarrow \mu \gamma$ or $(g-2)_\mu$. Among these constraints, the most stringent bound arises from $\tau \rightarrow \mu \gamma$ and translates into $\sqrt{|y_{\mu\tau}|^2+|y_{\tau\mu}|^2}< 0.016$ at 90\%C.L. for a sufficient SM-like Higgs. Other contributions from the other scalar particles should also be calculated. Assuming $y_{22}^{LLE} = y_{33}^{LLE} = 0$, then the only alternative way for $\tilde{l}^0$ to close the $\tau \mu$ transition moment loop is through the mass vertex with $h_d^0$. The formula for these contributions to the Wilson-coefficients $C_{L,R}$ are similar to those in Ref.~\cite{Harnik:2012pb, Blankenburg:2012ex}. Although the coupling constants are of the order $y_{23,32}^{LLE}$, $y_{33}^l$, which is much larger than the $y_{\mu \tau}$, $y_{\tau \mu}$, $y_{\tau \tau}$, the masses of the scalars $m_{h_d^0, \tilde{L}^0}^2$ are usually larger than the SM-like Higgs mass. The contributions are further suppressed by the cross mass term between  $\tilde{l}^0$ and $h_d^0$. As a result, they are usually of a similar order of magnitude as those contributed from the SM-like Higgs particle loops.  In addition, the loops involving the CP-odd Higgs particles can be neglected in the decoupling limit. Since the condition (\ref{effcoupval}) is an order of magnitude smaller than the bound from $\tau \rightarrow \mu \gamma$, our model can easily escape from this constraint. We can also adjust the signs and values of $y_{22}^{LLE}$ or $y_{33}^{LLE}$ in order to cancel out the remaining $C_{L,R}$ if one is still worried about the possible too large $\tau \rightarrow \mu\gamma$ branching ratio.


\section{Summary and Conclusions} \label{Conclusions}
In this paper, we have presented an R-parity violating supersymmetric model extended with $5+\overline{5}$ vectorlike particles that successfully addresses the flavor anomalies recently observed at the LHC within a unified framework. On the one hand, the combination of $U_iE_j\overline{D}$-type and $Q_i H_d D$-type RPV operators yields a NP contribution $C_9^{\mu,\mathrm{VLRPV}} = C_{10}^{\mu,\mathrm{VLRPV}} $, which is able to explain the anomalies in the $b\rightarrow s$ transition together with the MSSM box contributions. On the other hand, the SM-like Higgs obtains the LFV decay $h\rightarrow \mu\tau$ via the mixing with the sneutrino $\tilde{l}^0$. Both this mixing and the $\tilde{l}^0 \mu\tau$ vertices (given by the $L_i L E_j$ operators) arise naturally due to the RPV in our  model. All these explanations are compatible with various experimental measurements, especially the recent results of the $B_s \rightarrow \mu^+\mu^-$ decay. In our scenario, the mass of the VL down-type quark is of order TeV scale ($m_{D} \lesssim 800$ GeV in our benchmark point), which can be discovered or excluded by the future run of the LHC.

\section*{Acknowledgements:} 
We would like to thank Wolfgang Altmannshofer and David Straub for helpful correspondence. W.C.H thanks Qin Qin for useful conversations on flavor physics and Y.L.T thanks Chun Liu, Jiashu Lu, Shouhua Zhu and Da-Xin Zhang for discussions.  We are grateful to Jing Shu for a careful reading of the manuscript.

\appendix
\section{box contribution}\label{app_box}
Here we copy the loop function $f_9^{\mathrm{box}}$ entering the Wino boxes contributions from Appendix of Ref.~\cite{Altmannshofer:2013foa} and correct a typo in it
\begin{eqnarray}
f_9^{\mathrm{box}}(x,y) &=& \frac{12(x-2y+xy)}{(1-x)(y-x)(1-y)^2}-\frac{12x^2\log x}{(1-x)^2(x-y)^2}+\frac{12y(2x-y-y^2)\log y}{(x-y)^2(1-y)^3} \xrightarrow{x,y \rightarrow 1} 1 \notag \\
\end{eqnarray}

\section{The Perturbative diagonalizing of the squared mass matrix (\ref{MassMatrixBeforeRotation})}
In order to treat perturbatively  with as fewer and smaller non-diagonal elements as possible,
we first rotate the mass matrix (\ref{MassMatrixBeforeRotation}) into the Goldstone basis by
\begin{eqnarray}
V = \left[
\begin{array}{ccc}
\sin \beta & -\cos \beta & 0 \\
\cos \beta & \sin \beta & 0 \\
0 & 0 & 1
\end{array}
\right],
\end{eqnarray}
then $M^2_{S} = V^{\dagger} M^2_{\text{Higgs+$\tilde{L}$}} V$ reads
\small{
\begin{eqnarray}
M^2_{S} = \left[
\begin{array}{ccc}
4 v^2 \lambda_{\mathrm{eff}} \cos^2 2 \beta & 2 v^2 \lambda_{\mathrm{eff}} \sin 4 \beta & 0 \\
2 v^2 \lambda_{\mathrm{eff}} \sin4 \beta & -\csc \beta \sec \beta ( B \mu - 1.5 v^2 \lambda_{\mathrm{eff}} \sin 2 \beta + 0.5 v^2 \lambda_{\mathrm{eff}} \sin 6 \beta ) & -B_L \epsilon_L \sec \beta \\
0 & -B_L \epsilon_L \sec \beta & (M_{\tilde{L}}^2 + v^2 \lambda_{\mathrm{eff}} \cos 2 \beta)
\end{array}
\right]. \label{FirstRotation}
\end{eqnarray}}
Note that the element $M^2_{S,11}$ gives the same upper bound on the tree-level mass of the SM-like Higgs boson as in MSSM, so large loop contributions from the top squark or even an extension with one singlet field\cite{Hall:2011aa,Ellwanger:2009dp} is required to yield a 125 GeV Higgs mass. 
These contributions mainly modify the $M^2_{S,11}$ and thus have negligible effects on $V_l$, which is mostly determined by the lower right submatrix of (\ref{FirstRotation}). 
Here we focus on favor anomalies, we leave this Higgs mass issue aside. 

In the case of $(B_L\epsilon_L \sec \beta) /M_{\tilde{L}}^2 \ll 1$,  we can diagonalize the above matrix in terms of perturbative method and obtain at the lowest order 
\begin{eqnarray}
V_{u}^{\mathrm{lowest}}&=& 1 - 2\left( \frac{ \lambda_{\mathrm{eff}} v^2 \sin^2 2\beta \cos 2 \beta}{B \mu +  (\sin 6\beta - \sin 2\beta)\lambda_{\mathrm{eff}}v^2} \right)^2\notag \\
V_{d}^{\mathrm{lowest}}&=&  \frac{ \lambda_{\mathrm{eff}} v^2 \sin 2\beta \sin 4 \beta}{ B \mu +  (\sin 6\beta - \sin 2\beta)\lambda_{\mathrm{eff}}v^2} \notag \\
V_{l}^{\mathrm{lowest}} &=& \frac{2 B_L \epsilon_L \lambda_{\mathrm{eff}} v^2 \sin \beta \sin 4 \beta}{ M_{\tilde{L}}^2 (B \mu +  (\sin 6\beta - \sin 2\beta)\lambda_{\mathrm{eff}}v^2)}. \label{VlLowest}
\end{eqnarray}

\bibliography{flavor_anomalies}

\end{document}